\documentclass[onecolumn,final]{svjour2}
\usepackage{amstext}
\usepackage{amssymb}
\usepackage{graphicx}
\usepackage{amsmath}

\makeatletter
\@ifundefined{textcolor}{}
{%
 \definecolor{BLACK}{gray}{0}
 \definecolor{WHITE}{gray}{1}
 \definecolor{RED}{rgb}{1,0,0}
 \definecolor{GREEN}{rgb}{0,1,0}
 \definecolor{BLUE}{rgb}{0,0,1}
 \definecolor{CYAN}{cmyk}{1,0,0,0}
 \definecolor{MAGENTA}{cmyk}{0,1,0,0}
 \definecolor{YELLOW}{cmyk}{0,0,1,0}
 }

\makeatother

\usepackage[english]{babel}

\begin{document}

\title{Speeding up evolutionary search by small fitness fluctuations }

\author{Jakub Otwinowski \and Sorin~Tanase-Nicola \and Ilya Nemenman}

\institute{Jakub Otwinowski, \email{jotwinowski@physics.emory.edu} \\Sorin Tanase-Nicola, \email{sorintan@physics.emory.edu} \at Department of Physics, Emory University, Atlanta, GA 30322, USA \and Ilya Nemenman, \email{ilya.nemenman@emory.edu} \at Departments of Physics and Biology and Computational and Life Sciences Initiative, Emory University, Atlanta, GA 30322, USA}
\date{\today}
\maketitle

\begin{abstract} 
  We consider a fixed size population that undergoes an evolutionary
  adaptation in the weak mutuation rate limit, which we model as a
  biased Langevin process in the genotype space. We show analytically
  and numerically that, if the
  fitness landscape has a small highly epistatic (rough) and
  time-varying component, then the population genotype exhibits a high
  effective diffusion in the genotype space and is able to escape
  local fitness minima with a large probability. We argue that our principal
  finding that even very small time-dependent fluctuations of fitness
  can substantially speed up evolution is valid for a wide class of
  models. \end{abstract}

\section{Introduction}

Organisms adapt to their environment by sequential fixation of
beneficial mutations. This process is often visualized as motion of a
population (specified by multi-dimensional genomic variables
corresponding to the dominant genotype in the population) in the
fitness landscape, where the height of the landscape corresponds to
the reproductive fitness of an average individual in the population
\cite{Wright1932}. Fitness landscapes are believed to be rough with
many local maxima, and a population may get stuck in one, so that
every plausible mutation is deleterious. In such cases adaptation to a
global fitness maximum requires the fixation of deleterious mutations,
which is rare. Even when there is a path of only neutral or weakly
selective mutations to the global optimum, it may be difficult to find
it, and navigating such paths can be slow due to the low fixation
probability ($\approx 1/N$ for a population of $N$ individuals for a
neutral mutation~\cite{Kimura1962}).

It has been recognized that temporal fluctuations in the fitness
landscape can drive the population out of a local fitness maximum. For example, the
maximum of fitness at one time may be on a fitness slope at another time,
allowing the population to leave the area. Such
connections between the fluctuating selective pressure, the population
size, and the population-genetics dynamics have been studied extensively
\cite{Gillespie1994,Mustonen2008}, starting with the introduction of
the concept of adaptive topography by Wright \cite{Wright1932}. More
recently the evolutionary dynamics of density regulated populations in
fluctuating environments have been elucidated in more ecologically
realistic models \cite{Heckel1980,MacArthur2001,Lande2009}, bridging
the gap between the classical population dynamics
\cite{Namba1984,Cushing1986} and the population genetics models.

In a recent pioneering numerical evolution experiment
\cite{Kashtan2007}, these ideas were further developed to show that
certain type of fluctuating environmental pressures speed up evolution
many times in a particular model. However, it remains unknown to what
extent these results generalize. Is the speedup a general property?
How does it depend on the spatiotemporal structure of the fluctuating
environment? Can a population escape any local maximum? How does the
motion in the genotype space depend on time? Do fitness fluctuations
have to be dramatic, as in Ref.~\cite{Kashtan2007}, or can small
fluctuations still speed the evolution up?

In this article, we answer some of these questions in the context of a
model of evolutionary dynamics that is simple enough to allow a
thorough analytical and numerical treatment, but is at the same time
general enough so that at least some of our predictions hold for a
wide class of evolutionary models. We consider the limit of a weak
mutation rate, when the time scales are well separated. The time
between successive mutations is slower that the typical fixation time,
and the characteristic time scale of the fitness landscape changes is
the longest. Further, we assume a constant population size, so that
the evolutionary dynamics depends only on the relative fitness
differences between the genotypes. We consider adaptation in a highly
epistatic genotypic space, such that the evolution takes place on a
one dimensional pathway with large, local fitness differences. Under
these assumptions, we show that the evolutionary search can be sped up
substantially when only a small component of the fitness landscape
undergoes temporal variations.

\section{The model}

Our model of a fluctuating environment is based on an overdamped
Langevin particle in a potential. The position $x$ is some generalized
coordinate that describes the dominant genotype in the population, and
hence a change in $x$ is a fixation event. This genotype changes with
the velocity given by
\begin{equation}
  \frac{dx}{dt}=-\frac{1}{\gamma}\frac{\partial U(x,t)}{\partial
    x}+\eta\label{eq:eom},\end{equation}
where $U$ is the potential, $\gamma$ is the friction, $\eta$
is a white Gaussian noise with the variance $2D$, where $D$ is the
intrinsic diffusivity. Notice that in the usual physics language, the process
will minimize the potential so that $U$ is the negative fitness.

Motion to minimize $U$ represents fixation of beneficial mutations,
while Langevin noise allows low probability fixation of neutral and
deleterious mutations. The first phenomenon is called natural
selection/drift in evolutionary/physics languages, and the second is
unfortunately referred to as drift/diffusion, respectively. To avoid
confusion, in the remainder of the article we use the physics
terminology.

We write the potential as 
\begin{equation}
U(x,t) = U_0(x)+\Phi(x)S(t),
\label{potential}
\end{equation}
and we focus on the following range of parameters:
\begin{align}
&{\rm var}\,S(t)\sim 1,\\
&\max[\Phi(x)]-\min[\Phi(x)]\ll \max[U_0(x)]-\min[U_0(x)]
\end{align}
This models the emergence of novel functions in a population. Namely, the
fitness is largely independent of time, as described by $U_0$.
However, small temporal changes in fitness are allowed. For example,
acquiring an enzyme that can metabolize a certain chemical is
generally advantageous if the chemical is present, but
somewhat deleterious if it is absent, so that the investment into
production of the enzyme cannot recovered. We model this by
adding the small  fitness component $\Phi(x)$ that
fluctuates as $S(t)$, representing, for example, changes in the
availability of the metabolite due to seasonal or geological
variations. Finally, we choose to separate the global, almost non-epistatic,
fitness from the local, possibly highly-epistatic (but small) effects by
making the gradient of $U_0$ smaller than that of $\Phi$,
even though the scale of $\Phi$ itself is smaller than that of $U_0$.

\begin{figure}
\begin{centering}
\includegraphics[width=0.5\textwidth]{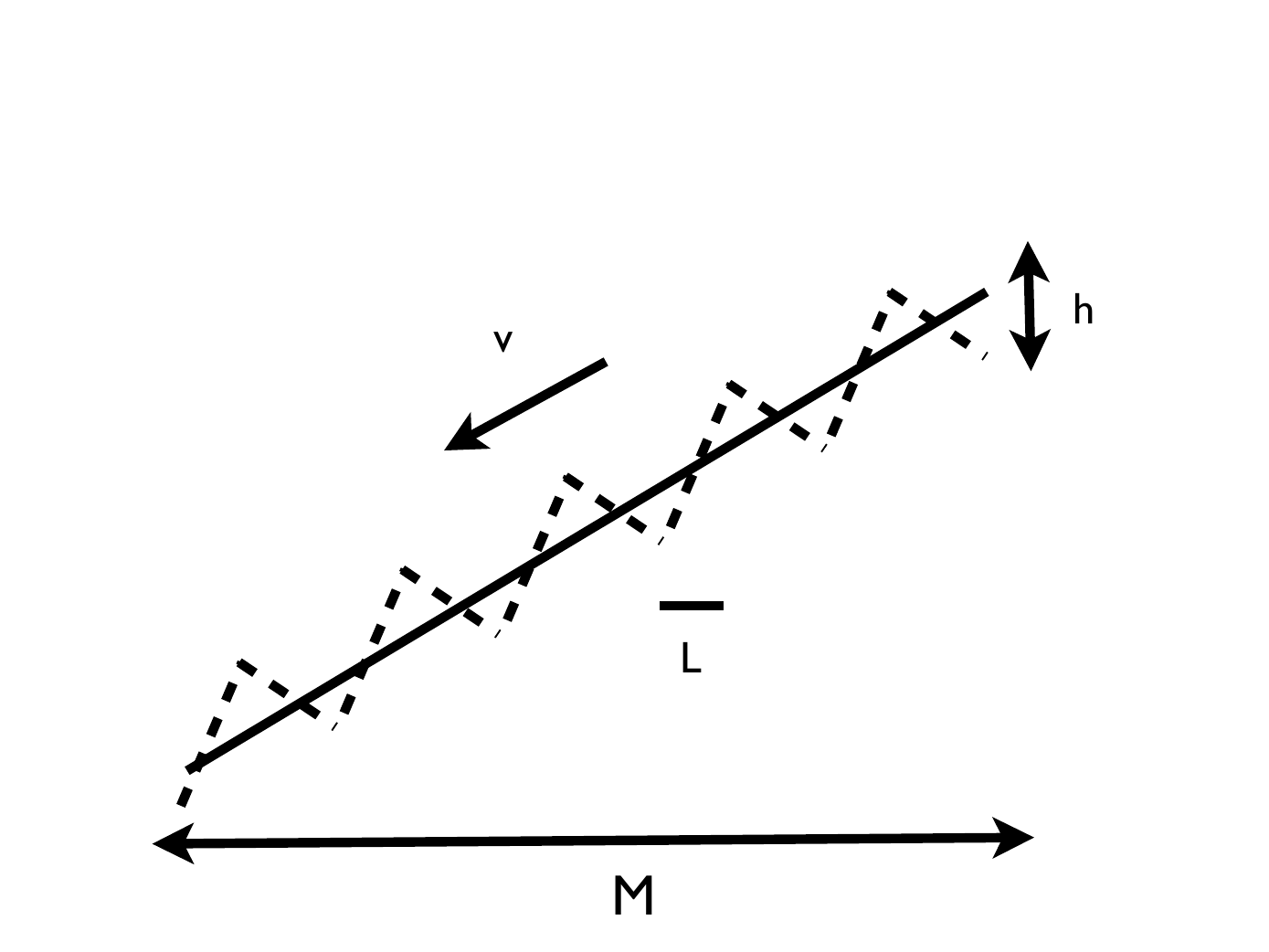}
\par\end{centering}
\caption{The potential $U(x,t)$ at a fixed time.  An oscillatory, symmetric, sawtooth perturbation is added on top of the average linear potential that creates a drift velocity of $v$. }
\label{fig:dia}
\end{figure}

With the conditions above, we can redefine $U_0$, $\Phi$, and $S$
without much loss of generality, so that $\langle S\rangle_t=0$. We
then consider the simplest form of $\Phi(x)$ and $S(t)$ that satisfy
these conditions, and we will discuss how our results generalize to
other forms of the functions in Section \ref{discussion}. Namely, we
choose $\Phi$ to be a zero-mean periodic saw-tooth potential, and $S$
to be a zero-mean periodic telegraph signal. These considerations
allow us to write near a particular point $x$ in the genotype space
\begin{align}
  &\frac{1}{\gamma}\frac{\partial U}{\partial x}=- v +\phi(x)s(t),\\
  &\phi(x)\equiv \frac{h}{L} \times{\rm sign}\left[\sin \frac{\pi x}{L}\right],\\
  &s(t)\equiv{\rm sign}\left[\sin \frac{\pi t}{T}\right],
\end{align}
where $v$ is the intrinsic drift or bias, defined as positive for the
drift to the right, see Fig.~\ref{fig:dia}. We always assume that
$|h|/L>|v|$, so that the fluctuating component of the potential can
actually create local maxima and minima on top of the global landscape
$U_0(x)$. In what follows we denote by $T$ the time between subsequent
potential flips (the half-period of the fluctuations), and $L$ is half
of the spatial period.

This model is similar to various stochastic ratchets considered in the
literature \cite{Magnasco1993,Tarlie1998,Dubkov2005,Sinitsyn2007}.
Thus, intuitively, the question of whether the fitness fluctuations
can speed up the evolutionary search is a question similar to whether
a rectified or a high-variance motion can appear due to ratcheting.

\subsection{\label{sec:Rescaling}Rescaling of the equation of motion}
Using  the choices above, we can rewrite the equation of motion,
Eq.~(\ref{eq:eom}) as
\begin{equation}
   \frac{dx}{dt}=\left[-\phi(x)s(t)+v\right]+\sqrt{2D}\eta.
   \label{eq:visual}
\end{equation}
where $\eta$ is a Gaussian white noise of unit variance. In
Eq.~(\ref{eq:visual}), the dynamics explicitly depends on five
different parameters $L$, $T$, $v$, $h$, and $D$. Nevertheless, by
rescaling the time, the space, and the potential as $x/L \to x$, $
t/T\to t$, $ \frac{L}{h}\phi\to \phi$, $\frac{L}{h}v\to v$, we can
reduce the number of parameters to only three: the ratio of the
typical diffusion time over half the spatial period to half of the
temporal period, $\omega=\frac{L^{2}}{2DT}$, the height of the
fluctuating barriers in diffusivity (temperature) units,
$\beta=\frac{h}{D}$, and the ratio between the slope of the average,
large scale potential to the slope of the fluctuating perturbation,
$v$. In physical terms, $\omega$ represents the diffusion time over
the distance $L$: if $\omega$ is large, the particle has time to
explore the entire valley of $\phi$ before the potential flips.
Further, $\beta$ measures the difficulty of crossing the peaks by
diffusion. Finally, the condition that the perturbation induces local
optima is $|v|<1$
 
 Using the rescaled variables, the dynamics becomes
\begin{equation} 
   \frac{dx}{dt}=\frac{\beta}{2\omega}\left[-\phi(x)s(t)+v\right]+\sqrt{\frac{1}{\omega}}\eta.\label{eq:sc}
\end{equation}
We will use these rescaled variables in the rest of the article,
unless noted otherwise. From this equation, it is easy to recover the
dynamics in the original, non-scaled units by simple multiplications.
In what follows we present simulation results obtained using first
order Euler integration scheme of the dynamics defined in rescaled
variables, Eq.~(\ref{eq:sc}).

\section{Fluctuating potentials enhances  diffusion and drift}

Numerical simulations suggest that the behavior of $x(t)$ at large
times is diffusive, and anomalous scaling is not seen
\cite{Weiss1986,Metzler2000,Lubelski2008,Bel2009}. We can characterize
the genotype coordinate motion by an effective drift and a diffusion
constant, which depend on the spatial and the temporal periods of the
fluctuations and the barrier height. To characterize the enhancement
or the suppression of the motion compared to the intrinsic diffusivity
and drift, we define 
\begin{align}
 r_{D}&\equiv\frac{D_{\textrm{eff}}}{D}=\frac{\textrm{var}_t(x)}{2t}\frac{L^{2}}{DT}=\frac{\textrm{var}_t(x)}{t}\omega\label{eq:rd},\\
r_{v}&\equiv\frac{v_{\textrm{eff}}}{v}=\frac{\langle
  x(t)\rangle}{t}\frac{L}{vT}=\frac{\langle
  x(t)\rangle}{t}\frac{2\omega}{v\beta} \label{eq:rv},
\end{align}
where the time-dependent means and variances of the trajectories
$x(t)$ are obtained numerically. As seen in the
Fig.~\ref{fig:Enhancement-of-diffusion}, both the effective drift and
the effective diffusion can be enhanced with respect to the intrinsic
values, this enhancement having a maximum for fluctuation periods
comparable to the average time to travel between two inflection
points.

\subsection{Building intuition}
When $\beta\ll1$, the sawtooth peaks are very small, and the diffusion
has no trouble crossing them. When $\beta$ is larger, the behavior is
more interesting. For $\omega\to0$, the particle has ample
time to fall into a minimum of $\phi$ before $s(t)$ flips. Then when
the potential flips, the particle can now go either left or right, both with
large probabilities, which creates a biased random walk behavior with
the effective diffusion coefficient $\approx{L^{2}}/({2T})$, or,
equivalently, $r_{D}\approx\omega$, and the system is only weakly
sensitive to the value of $v$. The fluctuating potential allows
the particle to diffuse against the drift, so that {\em the
speed of the evolutionary search  is
  strongly enhanced} when the environment oscillates.

\begin{figure} \begin{centering}
    \includegraphics[width=.5\textwidth]{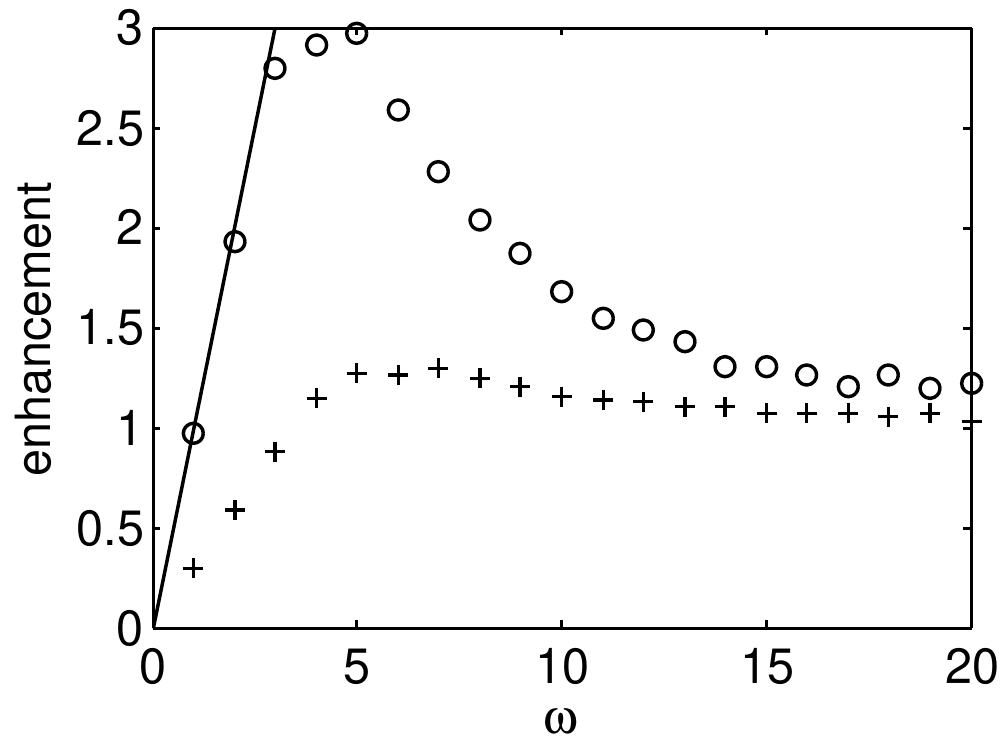}
    \par\end{centering}
  \caption{\label{fig:Enhancement-of-diffusion}Enhancement of
    diffusion, $r_{D}=\frac{D_{\rm eff}}{D}$ (circles) and drift,
    $r_{v}=\frac{v_{\textrm{eff}}}{v}$ (pluses) as a function of the
    relative flipping frequency $\omega=\frac{L^{2}}{2DT}$, with
    $\beta=10$ and $v=1/10$. At low $\omega$, the particle has time
    to reach the minima, and $D_{\rm eff}\approx\frac{L^{2}}{2T}$, or
    $r_{D}\approx\omega$. Simulations are averaged over 1000
    trajectories and 1000 time steps. Solid line indicates
    $r_D=\omega$. Error bars are smaller than the symbols.}
\end{figure}

For low $\omega$, $r_{D}\approx\omega$ is also small, so the diffusion
is suppressed compared to the internal value. However,
for $h\gg1$ and without the changing sign of $s(t)$, the particle
would get stuck at a minimum of $\phi$ almost immediately, and the
overall diffusion would be, essentially, zero: it is hard to exit a
deep potential only well with the help of diffusion. Whether $r_D$ is greater
or less than 1, the fact that oscillations make it nonzero is our most
important finding, suggesting that temporal fluctuations can make
fixation of rare-to-fixate mutations a much more common process.
Figure \ref{fig:Enhancement-of-diffusion} demonstrates these findings
for different values of  $\omega$ and for
$\beta=10$ and $v=1/10$. %

When the flipping is fast compared to the diffusion time (large
$\omega$ in the Figure), the fluctuating potential "blurs" together
and is effectively zero. The only motion is due to the internal
diffusion $D$ and $r_D$ goes to one.

\subsection{Analytical treatment at $\beta\to\infty$}

In the limit $\beta\gg1$, the peaks are very high, the particle almost
never crosses them due to noise, and analytical progress can be made.
First consider a particle that starts close to a local minimum $x_0$
of the sawtooth. After some time $\sim{D}/{v^{2}}$, which goes to
zero as $\beta\rightarrow\infty$, the particle equilibrates near $x_0$
with a probability density
 \begin{equation}
  p(x|x\gtrless x_0)\propto e^{-(1\mp v)|x|}.
  \end{equation}
  Then the ratio between the probability, $k_>$, that a particle is
  located to the right of $x_0$ and will move to the right after the
  potential flips to the probability, $k_<$, that it is to the left of
  $x_0$ and will move to the left is
  \begin{equation} \frac{k_>}{k_<}=\frac{1-v}{1+v}.\label{eq:k}
  \end{equation} When $s(t)$ changes sign, for a small $D$ the
  particle then glides down with a constant velocity in its chosen
  direction, reaching the next minimum to the right ($>$) or to the
  left ($<$) in \begin{equation}
  \tau_{\gtrless}=\frac{2\omega}{\beta}\frac{1}{(1\mp v)}.\label{eq:tp}
\end{equation} 
If $\tau_{\gtrless}<1$, then
the particle has the time to reach the minimum on either the left or
the right hand side (assuming, as always, that  the
sawtooth actually forms the local minima, i.e., $0<v<1$). Then when the
potential flips the next time, the process repeats. This results in a
discrete random walk between the extrema of the sawtooth, and (in
unscaled variables)
\begin{align}
  D_{\textrm{eff}}&=\left[(k_{>}+k_{<})-(k_{>}-k_{<})^{2}\right]\frac{L^{2}}{2T},\\
  v_{\textrm{eff}}&=\left(k_{>}-k_{<}\right)\frac{L}{T}. 
\end{align}
In dimensionless units, 
\begin{align}
  r_{D}&=\left[(k_{>}+k_{<})-(k_{>}-k_{<})^{2}\right]\omega,\\
  r_{v}&=\left(k_{>}-k_{<}\right)\frac{2\omega}{v\beta}.
\end{align}
Using Eq.~(\ref{eq:k}) results in: 
\begin{align}
  r_{D}&=\left(1-v^{2}\right)\omega\label{eq:rd1},\\
 r_{v}&=\frac{2\omega}{\beta}\label{eq:rv1}.
\end{align}
Notably $r_{D}\propto\frac{1}{D}\rightarrow\infty$ when $D\to0$, but $r_D/\beta=D_{\textrm{eff}}/h$ is finite. 

\begin{figure} \begin{centering}
    \includegraphics[width=0.5\textwidth]{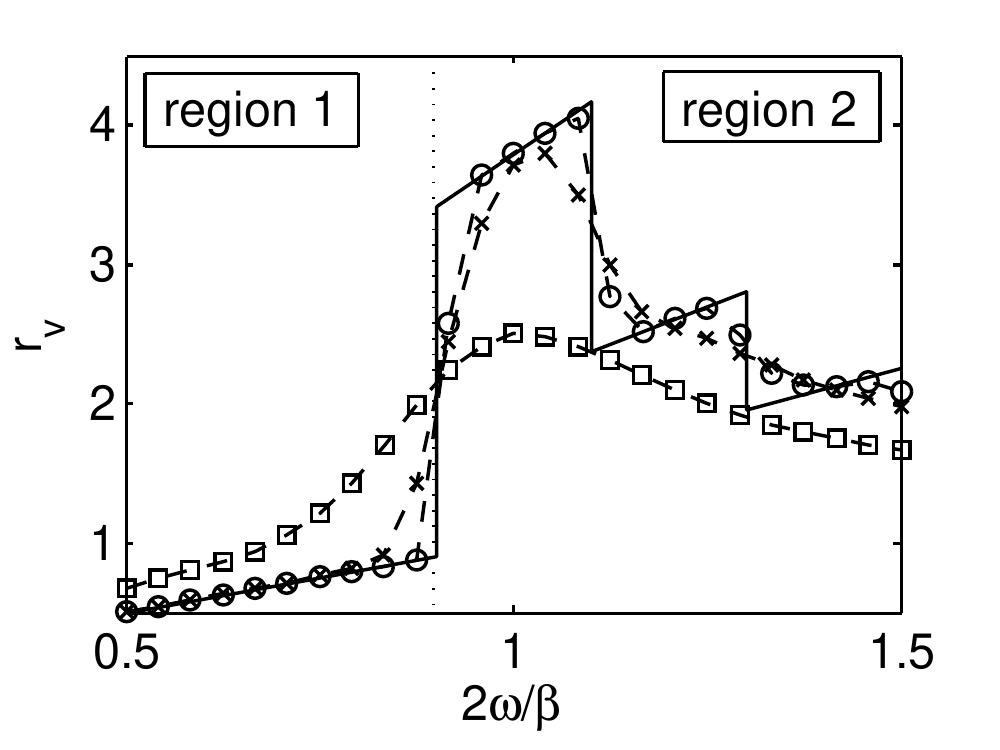}\includegraphics[width=0.5\textwidth]{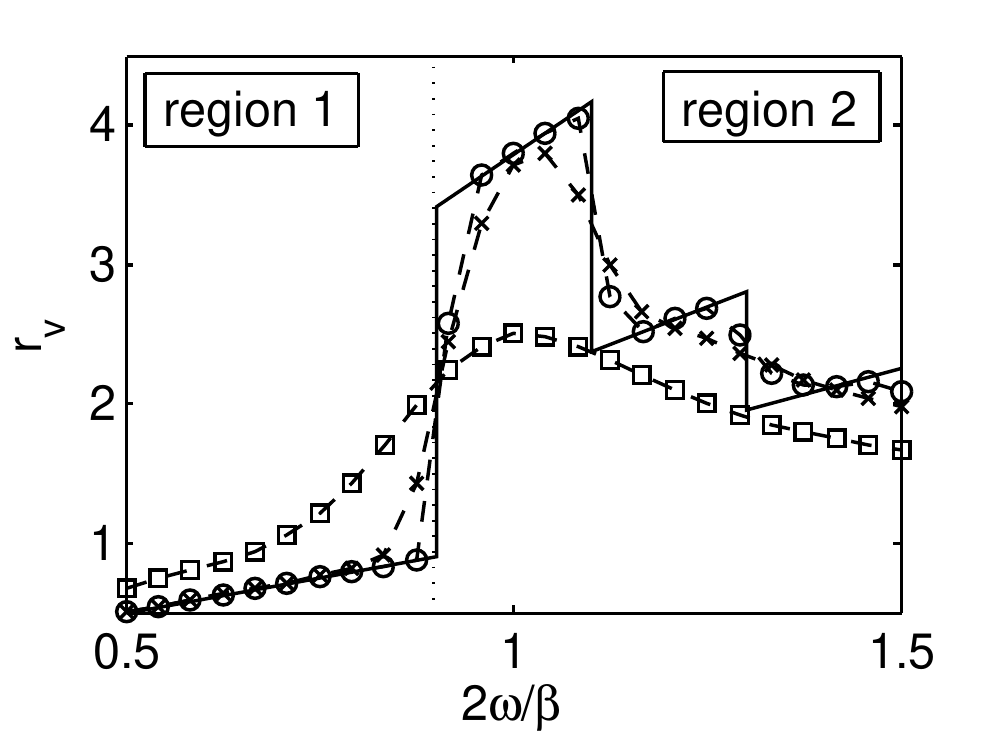}
    \par\end{centering} \caption{\label{fig:triangleplot}Effective
    diffusion (left) and effective drift (right) versus the period of
    the fluctuations. Notice that $r_D$ is normalized by $2/\beta$,
    and it remains finite even if $D\to0$ and $\beta\to\infty$. The
    data are obtained for (in decreasing order of noise strength)
    $\beta=100$ (squares), $\beta=1000$ (crosses), $\beta=10,000$
    (circles), $\beta\rightarrow\infty$ (solid line, analytical
    result). In the small noise case, the behavior of the diffusing
    particle is markedly different between Region 1 and Region 2.
    Region 1 corresponds to small $\omega$ when a particle can always
    travel between two extrema of the potential, performing an
    effective biased random walk. Region 2 corresponds to large
    $\omega$, when the particle spends most of the time traveling
    between minima but rarely reaching them. Simulations are averaged
    over 1000 trajectories and 1000 time steps.} \end{figure}

In Fig.~\ref{fig:triangleplot}, we compare the analytical results to
the numerically estimated $r_D$ and $r_v$ (here $r_{D}$ is normalized
by $\beta/2$). As $\beta\rightarrow\infty$, the
agreement is clearly seen for small $\omega$.

When the potential changes faster, and $\tau_{>}>1>\tau_{<}$, the
particle fails to make it to the minimum to the right and, after a
subsequent flip, always comes back to where it started from. However,
it always reaches the left minimum before the flip. When
$\tau_{>}>\tau_{<}>1$, it does not reach the left minimum either,
but goes the distance $\frac{\beta}{2\omega}(1+v)$ to the left, then
reverses and travels $\frac{\beta}{2\omega}(1-v)$ to the right, reverses
again and repeats until it reaches the left minimum. After one flip,
it moves $\frac{\beta}{2\omega}(1+v)$, after three flips it moves
$\frac{\beta}{2\omega}(1+3v)$, and so on. Eventually, when
$\beta/(2\omega)[(2n+1)v)+1]>1$, the particle reaches the next
minimum. Thus every time  $2\omega/(\beta v)$ crosses an odd  integer
more periods are needed to travel between the nearby  extrema, and the
diffusive behavior changes, resulting in the discontinuities  in
Fig.~\ref{fig:triangleplot}. This dynamics can be 
described by a master equation
\begin{equation}
P_{i}(t+1)=(1-k_{>})P_{i+1}(t-2n)+k_{>}P_{i}(t-1),
\label{master}
\end{equation}
where $P_i(t)$ stands for being at an extremum $i$ at the end of the
flip $t$, and exactly $2n+1$ flips are needed to travel between the
$i+1$'th and the $i$'th extrema. Solving this  for the drift and  the
diffusion (see Appendix) gives 
\begin{align}
r_{v}&=\frac{1-v}{3+v+2n(1-v)}\frac{2\omega}{v\beta},\\
r_{D}&=\frac{8(1-v^{2})}{(3+v+2n(1-v))^{3}}\,\omega.
\end{align}
The analytical results and the numerical simulations verifying them
are shown in Fig.~\ref{fig:triangleplot} with $\omega$ normalized by
$\beta/2$.

\section{Fluctuating potential shortens the fitness barrier crossing time}

We have shown that the typical diffusive/drift behavior of the system
is enhanced by the fluctuating component of the potential. However,
what kind of an effect does this enhancement have on the probability
of rare, atypical events, such as escape from a suboptimal fitness
maximum?

To model a barrier in fitness, we place the overdamped particle in a
flipping sawtooth potential and a constant force $v$, and we observe
the mean first passage time for the particle to reach an unscaled
distance $L\mathcal{M}$, against the drift, with a reflecting boundary
at $x=0$, see Fig.~\ref{fig:dia}.

\subsection{Fluctuation-activated escape from the minimum is possible even at zero internal diffusion}

\begin{figure}[t] \begin{centering}
    \includegraphics[width=0.5\textwidth]{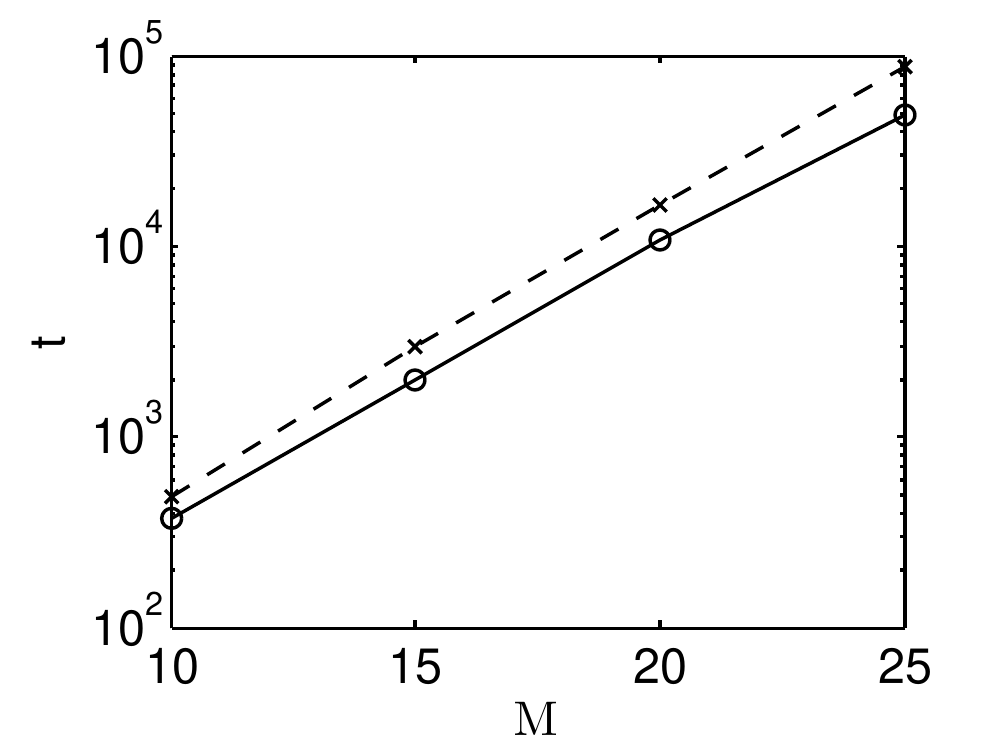} \par\end{centering}
  \caption{\label{fig:escape} Mean first passage times for crossing a
    barrier of width $\mathcal{M}$ in our model (simulation, circles)
    and in a continuous diffusion model with the appropriate effective
    parameters (crosses). The parameters are $\beta=10$, $\omega=2.5$,
    $v=-0.1$, and they correspond to the leftmost point in
    Fig.~\ref{fig:triangleplot}. The exit times from our model are
    faster than in the diffusion model. }
\end{figure}

If during one half period the particle is able to travel between the
two extrema, independent of the direction and without the help of the
noise, $\tau_{\gtrless}<1$, the probability that the particle travels
over multiple periods against the (effective) drift is finite even at
a very low noise. As seen in Fig.~\ref{fig:escape}, the escape time
over the average barrier in $U_0$ depends on the length of the barrier
$\mathcal{M}$. In order to understand this dependence, we consider our
model in the limit of zero fluctuating potential, which allows us to
use an analytical expression for the escape time \cite{Redner2001}
  \begin{equation}
\langle t\rangle_{D}=\frac{\mathcal{M}^{2}}{D}\left[\frac{1}{2Pe}-\frac{1}{4Pe^{2}}\left(1-e^{-2Pe}\right)\right],
\label{eq:esc}
\end{equation}
where $Pe$ is the P\'eclet number
 \begin{equation}
Pe=\left| \frac{vL{\mathcal{M}}}{2D}\right|.
\end{equation}

In Fig.~\ref{fig:escape}, we show the simulation results of the exit
time for slow fluctuations. We compare the results to what we would
have expected under a continuous diffusion with the drift and the
diffusion constant given by the values of the effective parameters as
obtained in the previous section. We observe that the escape times are
well approximated with the ``effective" continuous diffusion model.
Using a simple discrete space model does not improve the
approximation (not shown).

The conclusion is valid even for a very small intrinsic noise $D
\rightarrow 0$, which implies that the average time to escape over the
global barrier is fluctuation activated (as opposed to noise
activated), and is much faster.

\subsection{Fluctuations enhance escape even for steep barriers}

The qualitative behavior of particle trajectories changes if the
fluctuating potential is fast enough such that the particle can only
move less than one period in one direction in the absence of the
noise. Even though, on average, the variance of the particle position
grows linearly, and one can define a proper effective diffusion
coefficient, the particle never crosses a barrier against drift in the
absence of the intrinsic noise, $\beta\to\infty$. Hence the probability of rare
excursions against the effective drift cannot be described using the
same effective drift-diffusion model. 

\begin{figure}[t] \begin{centering}
    \includegraphics[width=0.5\textwidth]{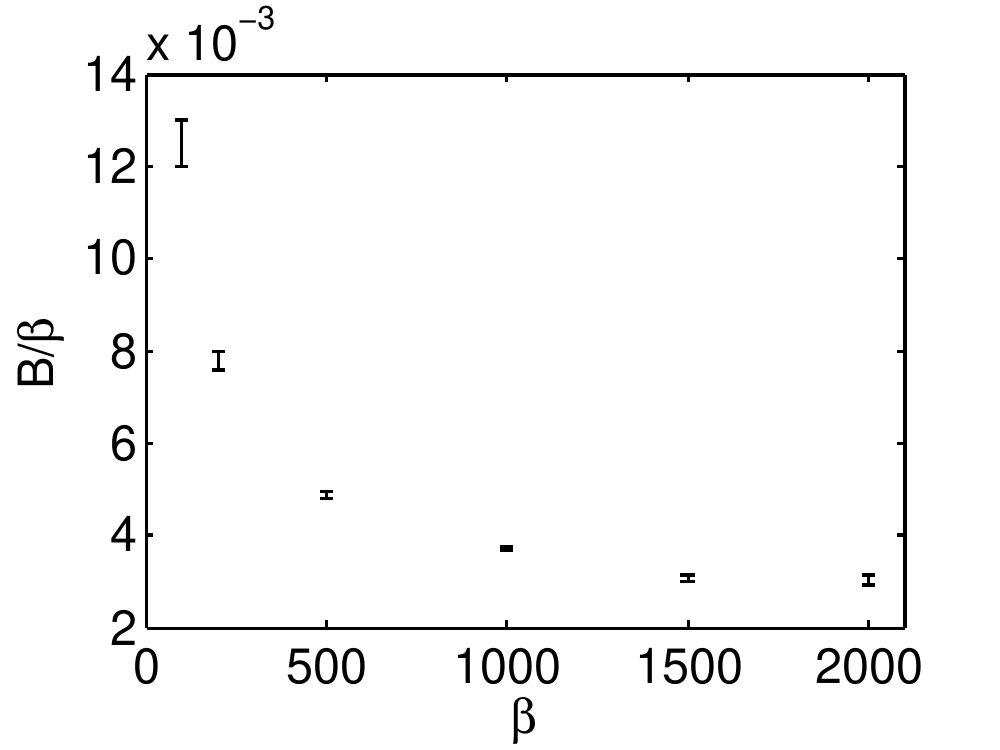}
    \par\end{centering} \caption{\label{fig:shortescape} Dependence of
    the effective Peclet number, $Pe_{\rm eff}=BM$ on the intrinsic
    noise. We simulate exit times for systems of lengths
    $\mathcal{M}=1,2,3,4,5$ for different $\beta$. For each dataset
    with the same $\beta$, we performed a weighted least squares fit
    for the average escape time of the form
    $t=A[\exp(B{\cal{M}})-1-B{\cal{M}}]$, fitting for $A$ and $B$. The
    error bars indicate the confidence bounds of the fits. We show
    $B/\beta$ versus $\beta$ decreases sublinearly and reaches a
    constant at $\beta\to\infty$. Thus the mean exit time goes to
    infinity, but it approaches it much slower than for diffusion with
    the fluctuating potential, for which $B/\beta=0.1$ for these
    parameter values, $2\omega/\beta=1$, $v=0.1$.}
\end{figure}

Fig.~\ref{fig:shortescape} reports results of numerical simulations at
different values of the additive noise. The escape time as a function
of $\mathcal{M}$ still can be fitted well with a "drift-diffusion"
model, Eq.~(\ref{eq:esc}), with a noise dependent effective P\'eclet number
$Pe(\beta)$. We conclude that, for small noises, the effective P\'eclet number
is approximatively inversely proportional to the noise strength (the
plot seems to reach a constant for $\beta\to\infty$). This implies
that the escape times will become infinite at zero noise. The
dependence is consistent with a model without any fluctuating
potential, but with some effective parameters. The parameters are such
that, with the fluctuations, the escapes are significantly faster.
Indeed as shown in Fig.~\ref{fig:shortescape}, the effective P\'eclet
number, defined as $Pe_{\rm eff}= B\mathcal{M}$, is always smaller
than the equivalent quantity in Eq.~(\ref{eq:esc}), such that
\begin{equation} B\ll\left| \frac{vL}{2D}\right|. \end{equation}

\section{\label{discussion}Discussion}

Using a one dimensional model of diffusion in the presence of a
constant force perturbed by small periodic fluctuations, we have shown
that time dependent potentials can significantly speed up the large
scale drift, diffusion, and barrier escape times. This conclusion is
valid even for a very small intrinsic diffusivity, the long time
statistics of the particle trajectories being mainly determined by the
properties of the potential fluctuations, but not of the Langevin
noise.

Our model is a caricature of evolutionary dynamics in the limit of low
mutation rates and constant population sizes. Even in this limit,
there are several simplifying assumptions in our model that can be
relaxed.

First, the periodicity of the potential time dependence is not
crucial. Based on the similarity with Brownian motor models
\cite{Magnasco1993,Tarlie1998}, we expect that our conclusions will
still be valid for a nonperiodic $s(t)$: the nonperiodic flipping will
mix together and average behaviors from the different regions in
Fig.~\ref{fig:triangleplot}. Further, Dubkov et al.~\cite{Dubkov2005}
studied a randomly flipping sawtooth with no drift. Their potential
flipped with dichotomous Markovian noise with rate $v$. They found an
analytic expression for the diffusion enhancement $r_{D}$: it grows
slower than in our model for small $\omega$, and the peak of $r_{D}$
is lower and at larger $\omega$. This is consistent with the averaging
over different regions in Fig.~\ref{fig:triangleplot}.

Our model is based on a piecewise linear periodic potential with
discontinuous first derivatives (sharp minima and maxima). We expect
that our conclusions stay valid as long as the spatially periodic
perturbations are sharp enough so that, upon a flip, a particle can
leave the vicinity of a maximum in a time much faster than a typical
travel time between the extrema. If the potential is not periodic,
this will introduce a quenched noise and is likely to result in
emergence of regions in $x$ that are very hard to cross, similar to
\cite{Weiss1986}.

If the assumption of well separated time scales and constant
population size are not satisfied, the Langevin microscopic dynamics
used here is not valid any more. Then different fitnesses can give
rise to different population sizes, and hence to varying fixation
rates and to a space dependent $D$ in our language, which would
require more complex models with position and time dependent noise
strengths. Such models would modify the probability of individual
trajectories \cite{Gopich2006,Bel2009}, and more work is needed in
order to identify the regimes in which such diffusion models and their
predictions apply to population genetics dynamics.

\appendix

\section{\label{sec:Diffusion-and-Drift}Diffusion and drift in the no-noise limit}

If it takes $2n+1$ flips to go from site $i+1$ to site $i$, and, going rightwards, one returns back in two flips, then
\begin{equation}
P_{i}(t+1)=(1-b)P_{i+1}(t-2n)+bP_{i}(t-1).\label{master:app}
\end{equation}
Multiplying by $i$ and summing over it, we get
\begin{equation}
\langle i (t+1)\rangle=(1-b)[\langle i(t-2n)\rangle+1]+b\langle i(t-1)\rangle.
\end{equation}
Assuming that the average can be written as
\begin{equation}
\langle i(t)\rangle=c+v_{\rm eff} t,
\end{equation}
we obtain
\begin{equation}
v_{\rm eff}=\frac{1-b}{1+2n(1-b)+b}.
\end{equation}
Now multiplying Eq.~(\ref{master:app}) by $i^2$ and summing over $i$, we get
\begin{equation}
\langle i^{2}(t+1)\rangle=(1-b)[\langle i^{2}(t-2n)\rangle+2\langle i(t-2n)\rangle+1]+b\langle i^{2}(t-1)\rangle.
\end{equation}
This allows to write for the variance of $i$ at moment $t+1$
\begin{multline}
\sigma^2(t+1)=(1-b)+(1-b)\sigma^{2}(t-2n)+b\sigma^{2}(t-1)+2(1-b)\langle i(t-2n)\rangle\\
-(1-b)[\langle i(t-2n)\rangle +\langle i(t+1)\rangle ]v(2n+1)\\
-2bv[\langle i(t-2n)\rangle+\langle i(t+1)\rangle].
\end{multline}
Now assuming
\begin{equation}
\sigma^{2}(t)=C+D_{\rm eff}t,
\end{equation}
we get
\begin{equation}
D_{\rm eff}=\frac{4(1-b)b}{\left(1+b+(1-b)2n\right)^3}.
\end{equation}

\section*{Acknowledgements}
We are grateful to S Boettcher, D Cutler, F Family, HGE Hentschel, B Levin, and
A Singh for stimulating discussions. We also thank J Lebowitz for
organization of the 102nd Statistical Mechanics Conference in
December 2009, which initiated our interest in the subject.

\end{document}